\documentclass[letterpaper,fleqn,usenatbib]{mnras}

\usepackage{newtxtext,newtxmath}

\usepackage[T1]{fontenc}
\usepackage{ae,aecompl}

\usepackage{booktabs}
\usepackage{longtable}
\usepackage{listings}
\usepackage{dcolumn}

\usepackage{soul}

\newcolumntype{K}{D{.}{.}{2,2}}
\newcolumntype{H}{D{,}{\pm}{4.4}}

\usepackage{graphicx}	

\title[Resolved centimetre-wave
  continuum in $\rho$~Oph~W]{Resolved spectral variations of the centimetre-wavelength continuum from the $\rho$~Oph~W photo-dissociation-region}

\author[Casassus et al.]{
  Simon  Casassus,$^{1}$\thanks{E-mail: simon@das.uchile.cl}
  Mat\'{i}as Vidal,$^{2}$
  Carla Arce-Tord,$^{1}$
  Clive Dickinson$^{3,4}$, \newauthor  
  Glenn J. White$^{5,6}$,  
  Michael Burton$^7$,
  Balthasar Indermuehle$^{8}$,
  Brandon Hensley $^{9}$, 
  \\
  $^{1}$  Departamento de Astronom\'{\i}a, Universidad de Chile, Camino El Observatorio 1515, Las Condes,  Santiago, Chile\\
  $^{2}$  Universidad Aut\'onoma de Chile, Facultad de
  Ingenier\'ia, N\'ucleo de Astroqu\'imica \& Astrof\'isica, Av. Pedro
  de Valdivia 425, Providencia, Santiago, Chile\\
  $^{3}$ Jodrell Bank Centre for Astrophysics, Alan Turing building, Department of Physics and Astronomy, School of Natural \\
  Sciences, The University of Manchester, Oxford Road, Manchester, M13 9PL, Manchester, U.K. \\
  $^{4}$ Cahill Centre for Astronomy and Astrophysics, California Institute of Technology, Pasadena, CA 91125, USA \\
  $^5$ RAL Space, Rutherford Appleton Laboratory, Chilton, Didcot, Oxfordshire, OX11 0QX, England\\
  $^6$ Department of Physics and Astronomy, The Open University, Walton Hall, Milton Keynes, MK7 6AA, England\\
  $^{7}$ School of Physics, University of New South Wales, Sydney NSW 2052, Australia\\
  $^{8}$ CSIRO Astronomy and Space Science, Marsfield NSW 2122, Australia\\
  $^{9}$ Spitzer Fellow, Department of Astrophysical Sciences, Princeton University, Princeton, NJ 08544, USA\\
 }

\date{Accepted XXX. Received YYY; in original form ZZZ}

\pubyear{2020}

\begin{document}
\label{firstpage}
\pagerange{\pageref{firstpage}--\pageref{lastpage}}
\maketitle

\begin{abstract}
  Cm-wavelength radio continuum emission in excess of free-free,
  synchrotron and Rayleigh-Jeans dust emission (excess microwave
  emission, EME), and often called `anomalous microwave emission', is
  bright in molecular cloud regions exposed to UV radiation, i.e. in
  photo-dissociation regions (PDRs). The EME correlates with IR dust
  emission on degree angular scales.  Resolved observations of
  well-studied PDRs are needed to compare the spectral variations of
  the cm-continuum with tracers of physical conditions and of the dust
  grain population. The EME is particularly bright in the regions of
  the $\rho$\,Ophiuchi molecular cloud ($\rho$\,Oph) that surround the
  earliest type star in the complex, HD\,147889, where the peak signal
  stems from the filament known as the $\rho$\,Oph-W PDR. Here we
  report on ATCA observations of $\rho$\,Oph-W that resolve the width
  of the filament.  We recover extended emission using a variant of
  non-parametric image synthesis performed in the sky plane.  The
  multi-frequency 17\,GHz to 39\,GHz mosaics reveal spectral
  variations in the cm-wavelength continuum.  At $\sim$30\,arcsec
  resolutions, the 17-20\,GHz intensities follow tightly the
  mid-IR, $I_\mathrm{cm} \propto I(8\,\mu$m), despite the breakdown
  of this correlation on larger scales. However, while the 33-39\,GHz
  filament is parallel to IRAC\,8\,$\mu$m, it is offset by
  15--20\,arcsec towards the UV source. Such morphological differences
  in frequency reflect spectral variations, which we quantify
  spectroscopically as a sharp and steepening high-frequency cutoff,
  interpreted in terms of the spinning dust emission mechanism as a
  minimum grain size $a_\mathrm{cutoff} \sim 6 \pm 1\,${\AA} that
  increases deeper into the PDR.
\end{abstract}



\begin{keywords}
radiation mechanisms: general --- radio continuum: general ISM ---
sub-millimetre -- ISM: clouds -- ISM: individual objects: $\rho$\,Oph
--- ISM: photodissociation region (PDR) -- ISM: dust
\end{keywords}



\section{Introduction}

Cosmic microwave background anisotropy experiments have identified an
anomalous diffuse foreground in the range of 10-90\,GHz
\citep{kog96,lei97}, which was confirmed, in particular, by the {\em
  WMAP} \citep[e.g.][]{Gold2011ApJS..192...15G} and {\em Planck}
missions \citep[e.g.][]{Planck2016A&A...594A...1P}. As summarised in
\citet{Dickinson2018NewAR..80....1D}, this diffuse emission is
correlated with the far-IR thermal emission from dust grains on large
angular scales, and at high galactic latitudes.  The spectral index in
specific intensity ($I_\nu \propto \nu^\alpha$) of the anomalous
Galactic foreground is $\alpha_\mathrm{radio/IR} \sim 0$ in the range
15-30\,GHz \citep{kog96}, but any semblance to optically thin
free-free is dissipated by a drop between 20-40\,GHz, with
$\alpha_\mathrm{radio/IR} \sim -0.85$ for high-latitude cirrus
\citep{dav06}. The observed absence of H$\alpha$ emission concomitant
to radio free-free emission would require an electron temperature $T_e
\geqslant 10^6$\,K to quench H\,{\sc i} recombination lines
\citep{lei97}.

The past couple of decades have seen the detection of a dozen
well-studied molecular clouds with bright cm-wavelength radiation in
excess of the expected levels for free-free, synchrotron or
Rayleigh-Jeans dust emission alone
\citep[e.g.][]{Finkbeiner2002ApJ...566..898F,
  Watson2005ApJ...624L..89W, Casassus2006ApJ...639..951C,
  Scaife2009MNRAS.394L..46A, Castellanos2011MNRAS.411.1137C,
  Scaife2010,Vidal2011MNRAS.414.2424V, Tibss2012ApJ...754...94T,
  Vidal2020MNRAS.495.1122V, Cepeda-Arroita2020arXiv200107159C}. A
common feature of all cm-bright clouds is that they host conspicuous
PDRs.  The {\em Planck} mission has also
picked up spectral variations in this excess microwave emission (EME)
from source to source along the Gould belt where the peak frequency is
$\nu_{\rm peak} \sim$26--30\,GHz while $\nu_{\rm peak} \sim 25\,$GHz
in the diffuse ISM \citep[][]{Planck2013A&A...557A..53P,
  Planck2016A&A...594A..10P}. The prevailing interpretation for EME,
also called anomalous microwave emission (AME), is electric-dipole
radiation from spinning very small grains, or `spinning dust'
\citep{dra98}. A comprehensive review of all-sky surveys and targeted
observations supports this spinning dust interpretation
\citep[][]{Dickinson2018NewAR..80....1D}. The carriers of spinning
dust remain to be identified, however, and could be PAHs
\citep[e.g.][]{Ali-Haimoud2014}, nano-silicates
\citep[][]{Hoang2016ApJ...824...18H, Hensley2017ApJ...836..179H}, or
spinning magnetic dipoles \cite[][]{HoangLazarian2016ApJ...821...91H,
  Hensley2017ApJ...836..179H}. A contribution to EME from the thermal
emission of magnetic dust
\citep[e.g.][]{DraineLazarian1999ApJ...512..740D} may be important in
some regions \citep[][]{DraineHensley2012ApJ...757..103D}.


Increasingly refined models of spinning dust emission reach similar
predictions, for given dust parameters and local physical conditions
\citep[e.g.][]{Ali-Haimoud2009, Hoang2010ApJ...715.1462H,
  Ysard2010A&A...509A..12Y, Silsbee2011MNRAS.411.2750S}. In addition,
thermochemical PDR models estimate the local physical conditions that
result from the transport of UV radiation
\citep[e.g.][]{LePetit2006ApJS..164..506L}. Therefore, observations of the
cm-wavelength continuum in PDRs can potentially calibrate the spinning
dust models and identify the dust carriers. The radio continua from
PDRs may eventually provide constraints on physical conditions.

In particular $\rho$\,Oph\,W, the region of the $\rho$\,Ophiuchi
molecular cloud exposed to UV radiation from HD\,147889, is among the
closest examples of photo-dissociation-regions (PDR), lying at a
distance of 138.9\,pc \citep[][]{Gaia2018yCat.1345....0G}. It is seen
edge-on and extends over $\sim$10$\times$3\,arcmin. $\rho$\,Oph is a region
of intermediate-mass star formation
\cite[][]{White2015MNRAS.447.1996W,Pattle2015MNRAS.450.1094P}.  It
does not host a conspicuous H\,{\sc ii} region, by contrast to the
Orion Bar, another well studied PDR, where UV fields are $\sim$100
times stronger. $\rho$\,Oph\,W has been extensively studied in the
far-IR atomic lines observed by ISO
\citep[][]{1999A&A...344..342L,Habart2003}. Observations of the bulk
molecular gas in $\rho$\,Oph, through $^{12}$CO(1-0) and
$^{13}$CO(1-0), are available from the COMPLETE database
\citep[][]{rid06}. While the HD\,147889 binary has the earliest
spectral types in the complex \citep[B2{\sc iv} and B3{\sc iv}
][]{cas08}, the region also hosts two other early-type stars: S\,1
\citep[which is a close binary including a B4{\sc v}
  star,][]{LadaWilking1984ApJ...287..610L}, and SR\,3 \citep[with
  spectral type B6{\sc v},][]{Elias1978ApJ...224..453E}. Both S\,1 and
SR\,3 are embedded in the molecular cloud. An image of the region
including the relative positions of these 3 early-type stars can be
found in \citet[][their Fig.\,4]{cas08} or in \citet[][their
  Fig.\,2]{Arce-Tord2020MNRAS.495.3482A}.

Cosmic Background Imager (CBI) observations showed that the
surprisingly bright cm-wavelength continuum from $\rho$\,Oph, with a
total {\em WMAP}\,33\,GHz flux density of $\sim$20\,Jy, peaks in
$\rho$\,Oph\,W \citep[][]{cas08}. The {\em WMAP} spectral energy
distribution (SED) is fit by spinning dust models \citep[within
  45\arcmin,][]{cas08}, as well as the {\em Planck} SED \citep[within
  60\arcmin,][]{Planck2011A&A...536A..20P}. However, the peak at all
IR wavelengths, i.e. the circumstellar nebula around S\,1, is
undetectable in the CBI data. Upper limits on the S\,1 flux density
and correlation tests with {\em Spitzer}-IRAC\,8$\mu$m rule out a
linear radio/IR relationship within the CBI 45\,arcmin primary beam
(which encompasses the bulk of $\rho$\,Oph by mass). This breakdown of
the radio-IR correlation in the $\rho$\,Ophiuchi complex is further
pronounced at finer angular resolutions, with observations from the
CBI\,2 upgrade to CBI \citep[][]{Arce-Tord2020MNRAS.495.3482A}.

Thus, while the cm-wavelength and near- to mid-IR signals in the
$\rho$\,Oph\,W filament correlate tightly, as expected for EME, this
correlation breaks down in the $\rho$\,Oph complex as a whole, when
including also the circumstellar nebula around S\,1.  Under the
spinning dust hypothesis, this breakdown points at environmental
factors that strongly impact the spinning dust emissivity per
nucleon. Dust emissivities in the near IR, from the stochastic heating
of very small grains (VSGs), are roughly proportional to the energy
density of UV radiation: $I_\mathrm{IR} \propto G_\circ
N_\mathrm{VSG}$. For a universally constant spinning dust emissivity
$\frac{j_{1\mathrm{cm}}}{n_H}$, we expect $I_\mathrm{cm} \propto
N_\mathrm{VSG} \propto I_\mathrm{IR} / G_\circ $. This correlation is
marginally ruled out in the CBI data, which thus point at emissivity
variations within the source \citep[][]{cas08}. The CBI\,2
observations, with a finer beam, reveal that
$\frac{j_{1\mathrm{cm}}}{n_H}$ varies by a factor of at least 26 at
3$\,\sigma$ \citep[][]{Arce-Tord2020MNRAS.495.3482A}.


Here we present observations of $\rho$\,Oph\,W acquired with the
Australia Telescope Compact Array (ATCA), with the aim of resolving
the structure of this PDR on $\sim$30\arcsec scales. The structure of
this article is as follows: Sec.\,\ref{sec:cabbobs} describes the ATCA
observations, Sec.\,\ref{sec:specvar} analyses the spectral variations
in these multi-frequency data, Sec.\,\ref{sec:crrls} reports limits on
the Carbon radio-recombination lines, which trace the ions possibly
responsible for the grain spin-up, and Sec.\,\ref{sec:conc}
concludes. Technical details on image reconstruction are given in the
Appendix.

\section{Observations} \label{sec:cabbobs}

\subsection{Calibration and imaging}



We covered the central region of $\rho$\,Oph\,W with Nyquist-sampled
mosaics, as detailed in the log of observations (see
Table\,\ref{table:log}). The primary calibrator was J1934-638 (for
flux and bandpass), and the secondary calibrator was J1622-253 (for
phase). The Compact Array Broadband Backend
\citep[CABB][]{CABB2011MNRAS.416..832W} provided a total of 4\,GHz
bandwidth split into two IFs. The data were calibrated using the
Miriad package \citep[][]{Miriad1995ASPC...77..433S} and following the
standard procedure for ATCA.


\begin{table*}
\centering
\caption{Log of observations for ATCA project C1845.}
\label{table:log}
\begin{tabular}{llllllll}
  \hline 
  Date         & Array$^{a}$          & Frequency$_1^b$ (beam)$^d$ & Frequency$_2^{c}$ (beam)$^d$  & Mosaic$^e$ \\  [0.05cm] \hline 
11-May-2009    &   H\,168  &  17481 (33.8$\times$25.9 / 85)   & 20160$^{c}$ (28.6$\times$21.4 / 271)  & 6 \\ 
12-May-2009    &   H\,168        &  5500 (46.1$\times$34.5 / 278)          & 8800 (28.3$\times$19.2 / 83)         & 1\\ 
08-Jul-2009    &   H\,75         &  33157 (17.2$\times$14.2 / 271 )        & 39157 (13.9$\times$11.3 / 276 )        & 6\\ \hline 
\end{tabular}
\begin{flushleft}
$^a$ ATCA array configuration. Antenna CA06, stationed on the North spur at $\sim$4\,km from the other 5 antennas in compact configuration, was not included in the analysis\\
$^b,^c$ Centre frequencies for the two CABB IFs. Each IF is made up of $2048\times 1$\,MHz channels \\
$^d$ Natural-weights beam in arcsec, in the form (BMAJ$\times$BMIN / BPA), where BMAJ and BMIN are the full-width major and minor axis, and BPA is the beam PA in degrees East of North. \\
$^e$ Number of fields in $\rho\,$Oph\,W.\\
\end{flushleft}
\end{table*}

Traditional image synthesis techniques based on the Clean algorithm
\citep[][]{Hogbom1974A&AS...15..417H} are best suited for compact
sources. Our initial trials with Miriad and Clean did not recover much
signal from the ATCA observations of $\rho$\,Oph~W, where the signal
fills the primary beam, and poor $uv$-coverage resulted in strong
negative sidelobes.  We thus designed a special purpose image
synthesis algorithm, which we call {\tt skymem}, based on
non-parametric model images rather than on the collection of
delta-functions used by Clean. Full details on image reconstruction
are given in Appendix\,\ref{sec:appendix}. {\tt skymem} yields
restored images defined in a similar way as for Clean.

Probably the most important feature of {\tt skymem}, which allowed the
recovery of the missing spatial frequencies, is the use of an image
template as initial condition. For adequate results this template must
  tightly correlate with the signal, and in this application we
used data from the Infrared Array Camera (IRAC) aboard {\em
  Spitzer}. The {\em c2d} Spitzer Legacy Survey provided an 8\,$\mu$m
mosaic of the entire $\rho$\,Oph region at an angular resolution of
2\,arcsec. The cm-wavelength radio signal in the $\rho$\,Oph\,W
filament is known to correlate with IRAC\,8\,$\mu$m
\citep[][]{cas08,Arce-Tord2020MNRAS.495.3482A}.  Since its resolution
is much finer than that of the ATCA signal we aim to image, and as it
is relatively less crowded by point sources compared to the shorter
wavelengths, we adopted the IRAC\,8\,$\mu$m mosaic as the {\tt skymem}
template after point-source subtraction by median filtering. This
template is the same as that used in \citep[][]{cas08}, and is also
shown here in the Appendix.

\begin{figure*}
  \begin{center}
    \includegraphics[width=\textwidth,height=!]{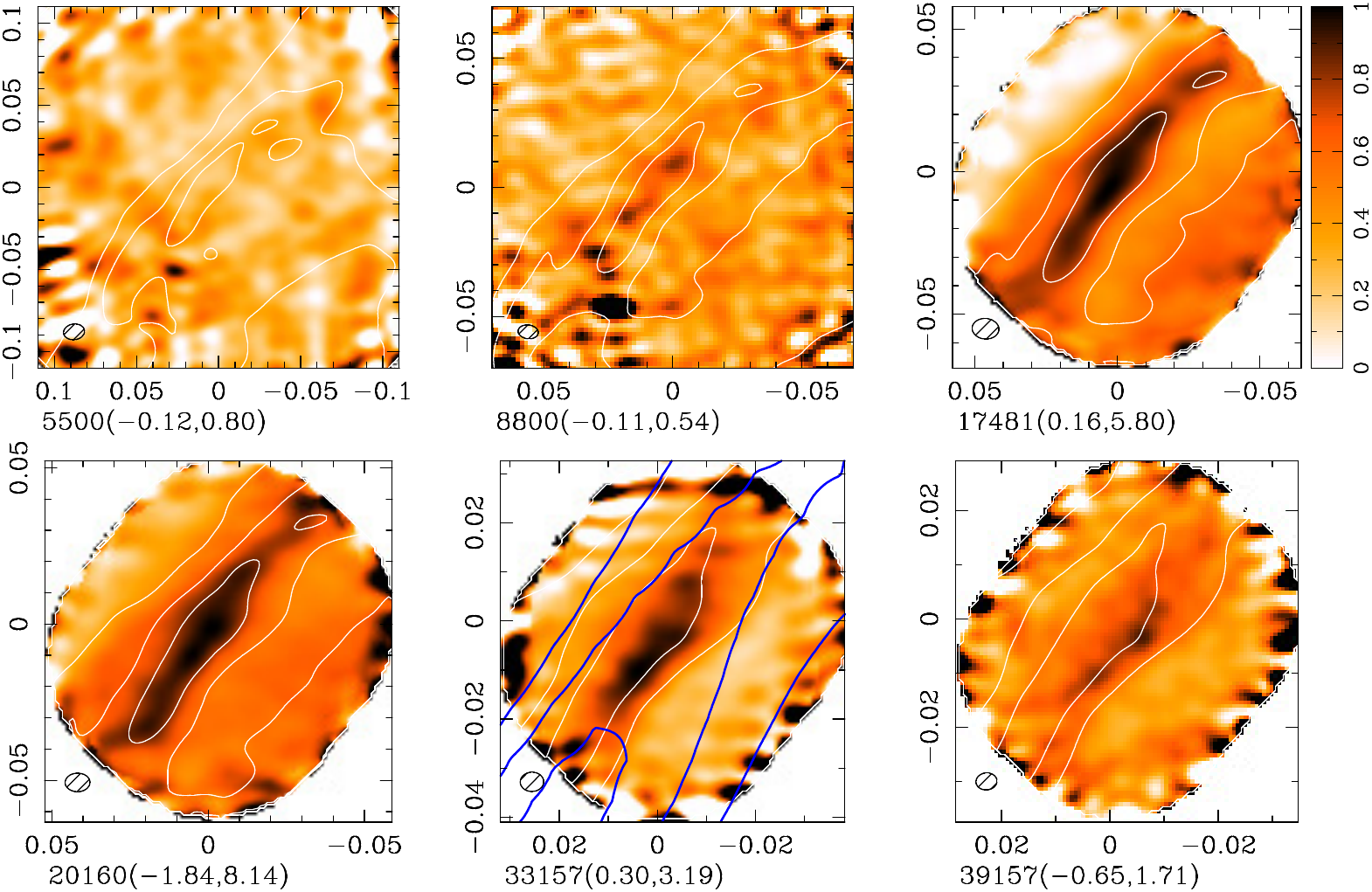}
  \end{center}
  \caption{\label{fig:ATCA} ATCA observations of $\rho$\,Oph\,W, and
    comparison with the 8\,$\mu$m emission, illustrating the
    systematic shift of the filament towards HD\,147889 (to the
    south-west) with increasing frequency.  $x-$ and $y-$axis show
    offset RA and DEC from $\rho$\,Oph\,W (J2000 16h25m57s,
    -24d20m50s), in degrees of arc. Note that fields of view are
    different as appropriate for each frequency.  The restored mosaics
    are shown in colour scale. Identical restorations of
    IRAC\,8\,$\mu$m visibilities, obtained by simulating the same ATCA
    observations at each frequency, are shown in white contours, with
    levels at 0.25, 0.5, and 0.75 times the peak IRAC\,8\,$\mu$m
    restored intensities.  All colour scales range linearly from 0 to
    1, and all images have been linearly scaled and normalised so that
    their entire intensity range, shown in parenthesis at the bottom
    of each image, is matched by the colour scale. Frequencies are
    given in MHz at the bottom left of each image. At 33.157\,GHz we
    have also overlaid the CBI2 {\tt uvmem} image of the 31\,GHz
    continuum from \citet{Arce-Tord2020MNRAS.495.3482A}, in blue
    contours at 0.5, 0.75, 0.95 times the peak. }
\end{figure*}

The {\tt skymem} mosaic for the ATCA data in $\rho$\,Oph\,W is
shown on Fig.\,\ref{fig:ATCA}. We can readily identify morphological
variations with frequency, so that the radio filament appears to
systematically shift towards the south-west, i.e. towards HD\,147889,
with increasing frequency.  There are however other frequency
dependent variations which may be due to interferometer filtering in
different $uv-$coverages. Since the morphology of the cm-wavelength
filament is similar to IRAC\,8\,$\mu$m, we compare the ATCA mosaics at
each frequency with {\tt skymem} reconstructions of IRAC\,8\,$\mu$m
after filtering for the corresponding $uv$-coverage. This accounts for
the spatial filtering by the interferometer and allows a robust
comparison between wavebands.

\subsection{Relevant point sources in the ATCA mosaics} \label{sec:PSs}

The ATCA angular resolutions allow to separate point sources from the
diffuse signal in the filament, such as proto-planetary disks whose
steeply rising thermal continuum emission may be relevant at the
higher frequencies. In particular, the disk around SR\,4, at J2000
16:25:56 -24:20:48.2 and near the center of coordinates in
Fig.\,\ref{fig:ATCA}, probably corresponds to the peak signal at
39\,GHz. It is best seen in the Clean map at 39\,GHz  shown in
Fig.\,\ref{fig:skymem}, since the extended signal is filtered in this
Clean image, and where we can infer a flux density of
0.34$\pm$0.03\,mJy. This point source has been subtracted from the
39\,GHz data shown in Fig.\,\ref{fig:ATCA} and  in the subsequent
analysis.

A different point source dominates the signal in the 5 and 8\,GHz
maps. This source coincides with the DoAr\,21 variable star, which is
surrounded by near-IR nebulosity and filaments
\citep[][]{Garufi2020A&A...633A..82G} but is not detected in the ALMA
continuum at 230\,GHz \citep[][]{Cieza2019MNRAS.482..698C}. In these
ATCA data, its flux density is 0.9$\pm$0.1mJy at 5\,GHz, and
4.2$\pm$0.05\,mJy at 8\,GHz. This point source is also picked up in
Clean images of the 17\,GHz and 20\,GHz ATCA data that include the
long baselines that join the 5 antennas in the compact configuration
with antenna CA06, stationed at $\sim$4.4\,km on the North spur of the
ATCA array. In these data the nebular emission is entirely filtered
out, and only DoAr\,21 remains, with flux densities of
0.6$\pm$0.03\,mJy at 17\,GHz and 0.2$\pm$0.04\,mJy at 20\,GHz. Antenna
CA06 is not included in the analysis of the nebular signal. DoAr\,21
is not subtracted in the analysis as it is located outside the field
of the higher frequencies, and its flux is negligible compared to the
nebular emission at 17\,GHz and 20\,GHz.


%

%
%

\subsection{ATCA - IRAC\,8$\mu$m cross-correlations} 
\label{sec:xcorr}


The signal in the ATCA reconstructions of $\rho$\,Oph\,W follows quite
tightly the IRAC filament, as also seen in other observations at
similar angular resolutions \cite[e.g. in LDN\,1246,  observed at
  25\arcsec by][using the Arcminute Microkelvin
  Imager]{Scaife2010}. Table\,\ref{table:xcorr} lists cross
correlation slopes and statistics.  The slopes $a(\nu)$ are calculated
on the {\tt skymem}-restored mosaics, with $I^R(\nu) =
a(\nu)I_{8\,\mu\mathrm{m}}$,
\begin{equation}
  a(\nu) = \frac{\sum_i I_{8\,\mu\mathrm{m}}(\nu,\vec{x}_i)
    I^R(\nu,\vec{x}_i) w_R(\nu,\vec{x}_i)}{\sum_i
    \left[I_{8\,\mu\mathrm{m}}(\nu,\vec{x}_i)\right]^2 w_R(\nu,\vec{x}_i)},
\end{equation}
where the weight image $w_R(\nu) = 1 / \sigma_R(\nu)^2$, and
$\sigma_R(\nu)$ is calculated with a linear mosaic of Miriad's
sensitivity maps for each pointing (see Appendix\,\ref{sec:appendix},
Eq.\,\ref{eq:noisemosaictheor}). The IRAC\,8\,$\mu$m comparison
images, $I_{8\,\mu\mathrm{m}}(\nu,\vec{x}_i)$, have been filtered for
the frequency-dependent $uv$-coverage, and scaled in intensity to
approximate the range of intensities observed in EME sources (as
described in Appendix\,\ref{sec:appendix}). Specifically, we scale the
IRAC\,8\,$\mu$m mosaic by the slope of the CBI/IRAC\,8\,$\mu$m
correlation measured in M\,78 by
\citet[][]{Castellanos2011MNRAS.411.1137C}. For instance, the ratio of
cm-wavelength specific intensities relative to IRAC\,8\,$\mu$m are
typically 3.05 times higher in $\rho$\,Oph\,W at 20\,GHz than in M\,78
at 31\,GHz. The slopes $a(\nu)$ are therefore dimensionless, and can
be used as an SED indicator, albeit in arbitrary units.

The linear-correlation coefficient $r_w$ in Table\,\ref{table:xcorr}
corresponds to
\begin{equation}
  r_w   =  \frac{\sum_i  (x_i - x_\circ) (y_i - y_\circ) }{\sqrt{\sum_i (x_i-x_\circ)^2 \,\sum_i(y_i-y_\circ)^2}}, \label{eq:rw}
\end{equation}
where $x_\circ = \frac{\sum_i (x_i \, w_i)}{\sum_i w_i}$, $y_\circ =
\frac{\sum_i (y_i \, w_i)}{\sum_i w_i}$, and
$x_i=I_{8\,\mu\mathrm{m}}(\nu,\vec{x}_i)$, $y_i =
I_\mathrm{ATCA}(\nu,\vec{x}_i)$. According to this correlation test,
the best match to IRAC\,8\,$\mu$m corresponds to 20\,GHz.



%
%

\begin{table}
\centering
\caption{ATCA - IRAC\,8\,$\mu$m correlation statistics}
\label{table:xcorr}
\begin{tabular}{lll}
\hline 
     Frequency$^a$ & $a$ $\pm \sigma(a)\,^b$  &  $r_w\,^c$    \\ \hline
      8800 &   0.13 $\pm$  0.01 &                0.52  \\   
     17481 &   1.63 $\pm$ 0.01  &                0.89  \\   
     20160 &   3.05 $\pm$ 0.02  &                0.95  \\   
     33157 &   2.77 $\pm$ 0.05  &                0.83  \\   
     39157 &   1.38 $\pm$ 0.04  &                0.61  \\ \hline
\end{tabular}
\begin{flushleft}
  $^a$ Centre frequency in MHz. 
  $^b$ dimensionless correlation slope $I_\mathrm{ATCA} = a  I_{{\rm IRAC}\,8\,\mu\mathrm{m}}$.
  $^c$ linear-correlation coefficient.

\end{flushleft}
\end{table}

\section{Spectral variations } \label{sec:specvar}

\subsection{Morphological trends with frequency}

\begin{figure*}
  \begin{center}
    \includegraphics[width=\textwidth,height=!]{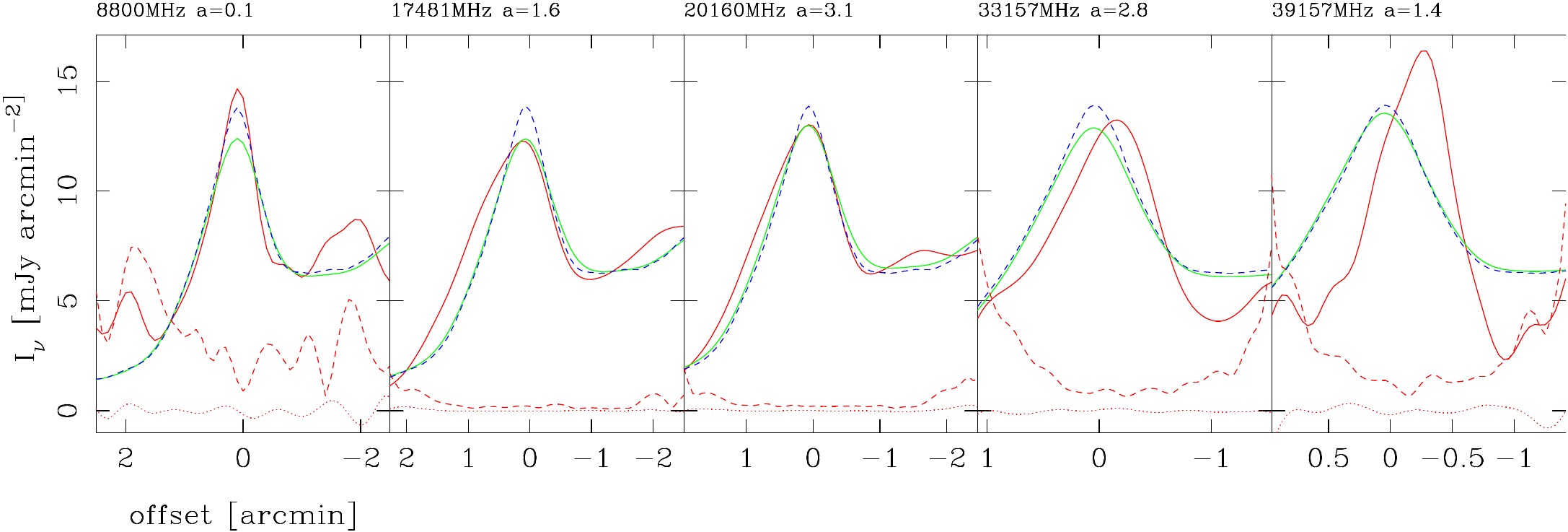}
  \end{center}
  \caption{\label{fig:prof} Profiles from the restored ATCA mosaics,
    extracted perpendicularly to the $\rho$\,Oph\,W PDR.  Frequencies
    are indicated on the top left.  The following number is the
    radio-IR correlation slope $a$ (same as in
    Table\,\ref{table:xcorr} and described in Sec.\,\ref{sec:xcorr}).
    The profile at each frequency is divided by $a$.  The $x-$axis
    shows offset in arcmin from the reference position, at J2000
    16:25:57.984 -24:20:37.760. $y-$axis shows specific intensities
    averaged in a region $\pm$1\,arcmin along the filament. The ATCA
    profiles are shown in {\bf red}, with the restored image in solid
    line, its average residuals in dotted line, and the rms-dispersion
    of residuals in dashed line.  The {\bf blue} dashed line is the
    average profile of the IRAC\,8$\mu$m template, which is the
    original IRAC image filtered and scaled by a reference radio/IR
    correlation slop (see Sec.\,\ref{sec:xcorr}). The {\bf green}
    dashed line is the corresponding profile of a simulation of the
    ATCA observations and {\tt skymem} restoration. }
\end{figure*}

The morphological variations with frequency apparent in
Fig.\,\ref{fig:ATCA} can also be measured with intensity profiles
across the $\rho$\,Oph\,W filament. The intensity profiles shown on
Fig.\,\ref{fig:prof} were generated by extracting 2\,arcmin-wide cuts
orthogonal to the filament. The images were rotated so that their
$y$-axis is aligned at a position angle of --40\,deg East of North,
and roughly coincident with the direction of the filament.
One-dimensional profiles were then obtained by averaging the 2-D
specific intensities along the $y-$axis.

%

\begin{figure}
  \begin{center}
     \includegraphics[angle=0,width=\columnwidth]{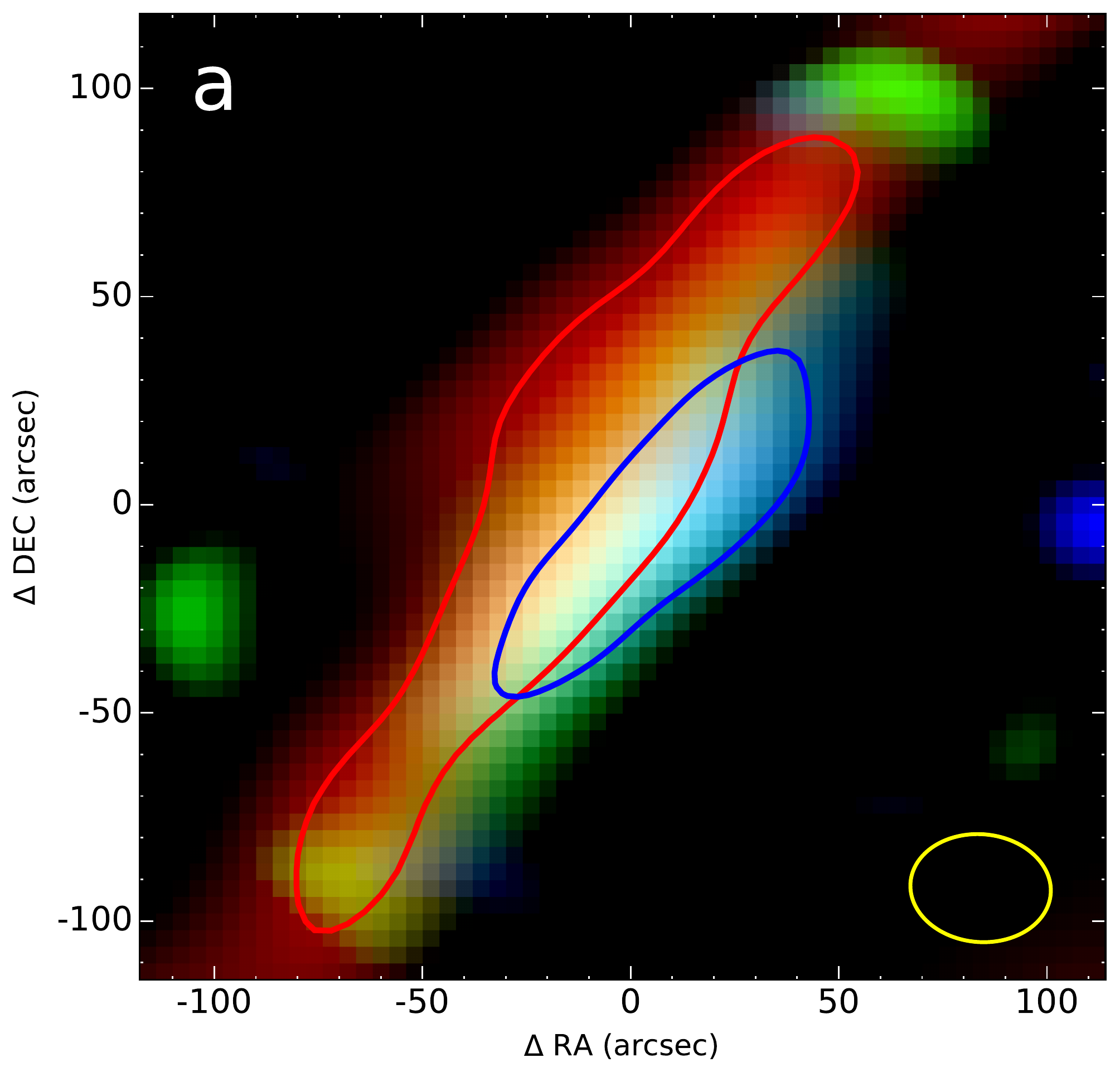}
     \includegraphics[angle=0,width=\columnwidth]{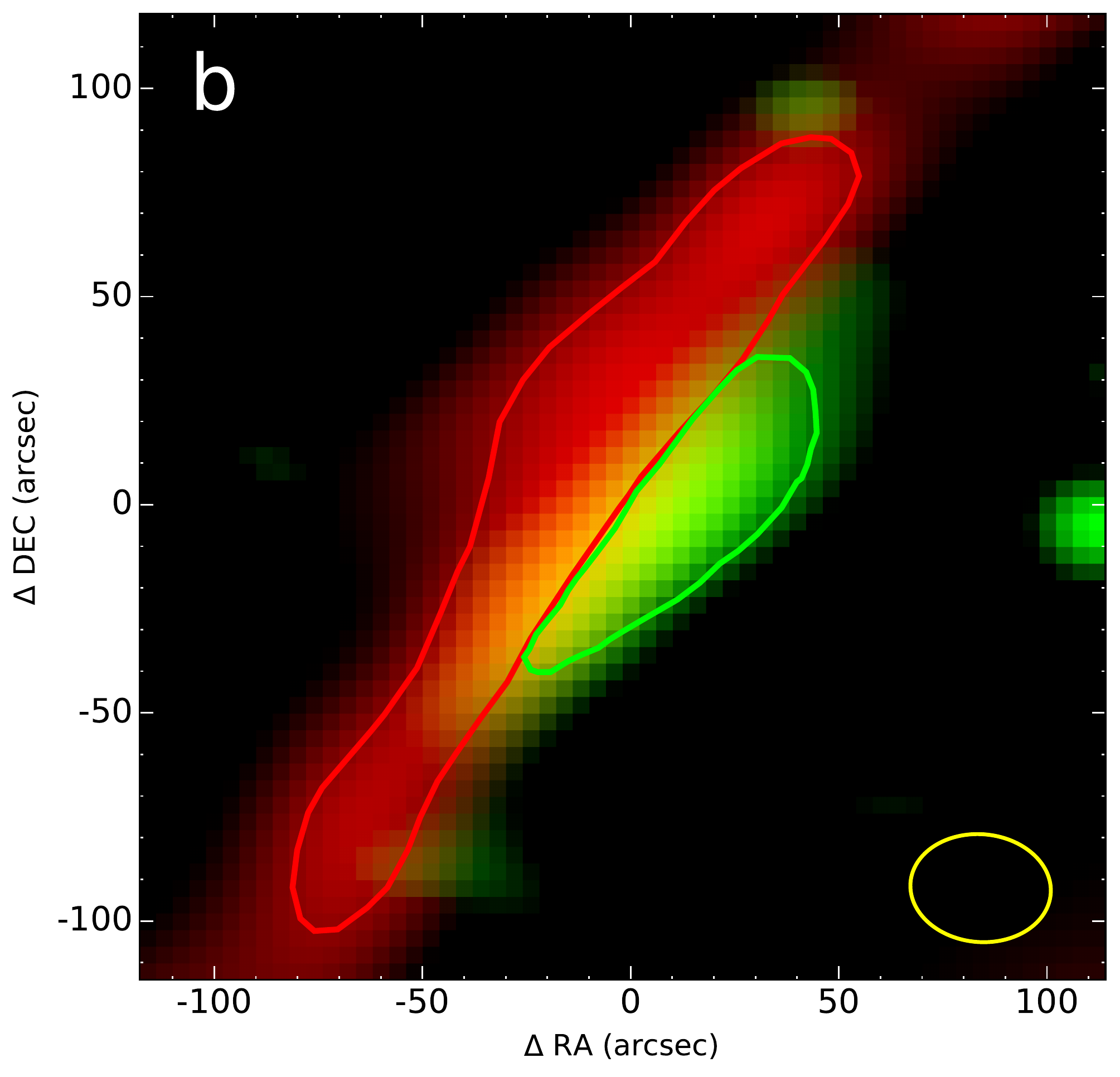}
  \end{center}
  \caption{ The $\rho$\,Oph\,W filament shifts towards the exciting
    star with increasing frequency, as illustrated in these
    colour-coded versions of the restored maps shown in
    Fig.\,\ref{fig:ATCA}, after subtraction of the SR\,4 point source
    at 39\,GHz, and degraded to a common angular resolution. {\bf
      a:} This RGB image is linearly scaled to cover the whole range
    of intensities at each frequency: 17\,GHz in red and 33\,GHz in
    green and 39\,GHz in blue. The contours are taken at 90\% peak
      intensity, with matching colours.  The common beam corresponds
      to that of the 17\,GHz map, and is indicated by the yellow
      ellipse (see Table\,\ref{table:log}).  {\bf b:} In this
      Red-Green version, with 17\,GHz in red and 39\,GHz in green, we
      illustrate the photometric apertures used to measure the SED
      (Sec.\,\ref{sec:sed}). The contours correspond to the 85\%-peak
      level at 17\,GHz (in red), and 80\% at 39\,GHz, (in green), but
      are modified to avoid overlap.}
  \label{fig:rgb_mask}
\end{figure}

A dual-frequency comparison of the maps degraded to the same angular
resolution is shown in Fig.\,\ref{fig:rgb_mask}. The common beam is
that of the 17\,GHz measurements (see Table\,\ref{table:log}). The
contours in Fig.\,\ref{fig:rgb_mask}b compare the morphologies at 17\,GHz and 39\,GHz and
correspond to the two photometric apertures used for the extraction of
the SED, they are thus edited from genuine contour levels  to avoid
overlap. The contours also illustrate the westward shift of the peak
emission in frequency.

We conclude that the cm-wavelength signal from the $\rho$\,Oph\,W
filament shifts towards higher frequencies with decreasing distance to
the exciting star HD\,147889. In other words, these morphological
trends with frequency point at spectral variations in the EME spectrum
when emerging from the PDR towards the UV source. It is interesting to
note that a similar spectral trend has been reported in the PDR
surrounding the $\lambda$\,Ori region, where the EME signal peaks at
increasingly higher frequencies  towards the
UV source \citep[][]{Cepeda-Arroita2020arXiv200107159C}. The next
Section addresses how the spectral trends implicit in the morphological
trends observed in $\rho$\,Oph\,W could be related to varying physical
conditions under the spinning dust hypothesis.

\subsection{Spectral energy distribution} 
\label{sec:sed}

The multi-frequency radio maps of $\rho$\,Oph\,W allow for estimates
of its SED between 5 and 39\,GHz. The morphological trends in
frequency should be reflected in variations of the SED between the
emission originating around the 17\,GHz peak and the emission coming
from the vicinity of the 39\,GHz peak.  Such multi-frequency analysis
requires smoothing the data to a common beam, which is that of the
coarsest observations, at 17\,GHz. Once smoothed, we measured the mean
intensity in each map inside the two masks shown in
Fig. \ref{fig:rgb_mask}. Interferometer data are known to be affected
by flux-loss, i.e. missing flux from large angular scales not sampled
by the $uv-$coverage of the interferometer.  By using a prior image
not affected by flux-loss, i.e. as defined in
Appendix\,\ref{sec:appendix}, the {\tt skymem} algorithm allows to
recover such flux loss under the assumption of linear correlation with
the prior. Our simulations using the prior image recovered the missing
flux exactly, but since the cm-wavelength signal does not exactly
follow the prior, biases in the flux-loss correction scheme may affect
the SEDs reported here. We expect such biases to be small and include
them in the absolute calibration error of 10\%, given the tight
correlation with the near-IR tracers and especially with the
IRAC\,8\,$\mu$m image used to build the prior image.

\begin{figure}
  \begin{center}
     \includegraphics[angle=90,width=0.49\textwidth]{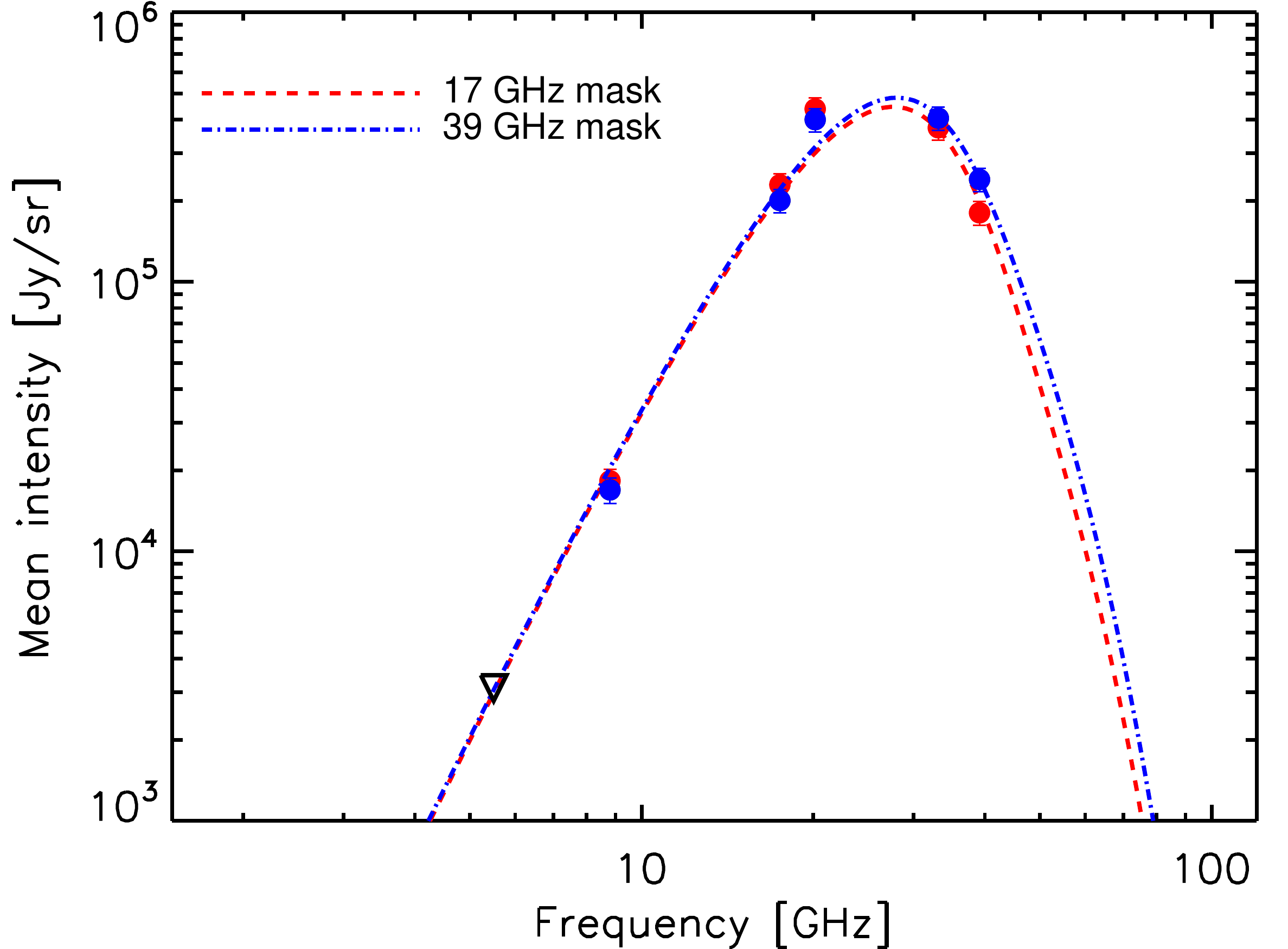}
  \end{center}
  \caption{SED of the $\rho$\,Oph\,W filament measured between 5.5 and
    39\,GHz within the two apertures shown in
    Fig. \ref{fig:rgb_mask}. The red and blue lines correspond to the
    best fit spinning dust models.}
  \label{fig:sed}
\end{figure}

Table\,\ref{tab:intensities} lists the mean intensities $\langle I_\nu
\rangle$ measured within the 17 and 39\,GHz masks, $\mathcal{M}_{17}$
and $\mathcal{M}_{39}$, for the six frequencies that we observed. We
weighted the photometric extraction using the noise image
$\sigma_R(\vec{x})$ given in Eq.\,\ref{eq:noisemosaic}, i.e.
\begin{equation}
  \langle I^k_\nu \rangle = \frac{\sum_{\vec{x}_j \in \mathcal{M}_k} w_R(\vec{x}_j) I_\nu(\vec{x}_j)}{\sum_{\vec{x}_j \in \mathcal{M}_k} w_R(\vec{x}_j) },
\end{equation}
for each mask $\mathcal{M}_k$, and with  $w_R = 1/\sigma^2_R$. The associated error is
\begin{equation}
\sigma(I^k_\nu) = \sqrt{ \frac{ N_{\rm beam} }{\sum_{\vec{x}_j \in \mathcal{M}_k} w_R(\vec{x}_j) }},
\end{equation}
where $N_{\rm beam}$ is the number of pixels in a beam.  The same SEDs
are also plotted in Fig.\,\ref{fig:sed}, where we have  included a
conservative 10\% systematic uncertainty.

When including the 10\% absolute flux calibration uncertainty, the
difference between the SEDs extracted in the two photometric apertures
is not significant. The $\chi^2$ distribution yields that the two SEDs
are different at 75\% confidence. Only the 39\,GHz average intensities
appear to differ at the 95\% confidence level, or
2$\sigma$. Table\,\ref{tab:intensities} nonetheless lists the ratio
between the measured intensities in each region
I$^{39}_{\nu}/$I$^{17}_{\nu}$, as this ratio systematically increases
with frequency, which may reflect the morphological trends.  In
Fig.\,\ref{fig:sed}, the spectra of the two regions show a steep drop
after reaching the peak, at a frequency of $\sim 30$\,GHz.  We can
notice that the difference in measured intensity between the two
regions is largest at 39\,GHz. The emission from the 39\,GHz mask
shows a spectrum brighter at higher frequencies than the emission
coming from the 17\,GHz mask. This is interesting as the 39\,GHz mask
is shifted towards the direction of the illuminating star HD\,147889.

\begin{table}
\centering
\caption{ATCA mean intensities and their ratio measured in two masks
  shown in Fig. \ref{fig:rgb_mask}. The quoted uncertainties refer to
  the thermal errors only - the actual uncertainties should include
  10\% in quadrature, and have been applied to the ratio $
  \rm{I}^{39}_{\nu} / \rm{I}^{17}_{\nu} $.}
\begin{tabular}{cHHH}
\toprule
    \multicolumn{1}{c}{Frequency}
   & \multicolumn{1}{c}{17\,GHz mask} 
   & \multicolumn{1}{c}{39\,GHz mask}
   & \multicolumn{1}{c}{Ratio} \\
    \multicolumn{1}{c}{[GHz]}
   & \multicolumn{1}{c}{[$\times 10^{4}$\,Jy/sr]}
   & \multicolumn{1}{c}{[$\times 10^{4}$\,Jy/sr]}
   & \multicolumn{1}{c}{$ \rm{I}^{39}_{\nu} / \rm{I}^{17}_{\nu} $} \\
     \midrule
 5.5   &    \multicolumn{1}{c}{$ <      0.3^{a}$} &  \multicolumn{1}{c}{$ <      0.3^{a}$} &  \multicolumn{1}{c}{$-$}   \\
  8.8 &     1.83 ,     0.04 &     1.69 ,     0.07 &      0.9 ,      0.1 \\
 17.5 &    22.93 ,     0.03 &    20.04 ,     0.04 &      0.9 ,      0.1 \\
 20.2 &    43.79 ,     0.06 &    39.99 ,     0.08 &      0.9 ,      0.1 \\
 33.2 &    37.32 ,     0.13 &    40.52 ,     0.19 &      1.1 ,      0.1 \\
 39.2 &    18.05 ,     0.08 &    23.99 ,     0.11 &      1.3 ,      0.2 \\
\bottomrule
 \end{tabular}
 \begin{flushleft}
   $^a 3\sigma$ upper limits using the dispersion of residuals.\\
\end{flushleft}
\label{tab:intensities}
\end{table}

\subsubsection{SED modeling} 
\label{sec:sed_modelling}

Previous works have shown that the cm-wave emission from this region
is dominated by EME and does not have major contributions from
synchrotron or free-free emission, and that its SED on degree angular
scales is adequately fit by spinning-dust models \citep[see
  e.g.][]{cas08,Planck2011A&A...536A..20P,Arce-Tord2020MNRAS.495.3482A}.
Here we ask the question of what are the consequences of the SED that
we measure with ATCA, on arc-minute scales, for the physical
conditions and grain populations within the cloud, and under the
spinning-dust hypothesis. Since the ATCA mosaic of $\rho$\,Oph\,W is
clear of any detectable free-free emission, as shown by the 5\,GHz
map, we used only a spinning dust component, as calculated using the
{\tt SPDUST} code \citep{Ali-Haimoud2009}.

The spinning dust emission depends on a large ($\sim$\,10) number of
parameters that determine environmental properties: the gas density
(n$_{\rm H}$), the gas temperature (T), the intensity of the radiation
field (parameterised in terms of the starlight intensity relative to
the average starlight background, $\chi$), the ionized hydrogen
fractional abundance $x_{\rm{H}} \equiv n_{\rm{H^+}}/n_{\rm{H}}$, the
ionized carbon fractional abundance $x_{\rm{C}} \equiv
n_{\rm{C^+}}/n_{\rm{H}}$. In addition, the spinning dust emissivities
also depend on the grain micro-physics, such as the grain size
distribution and the average dipole moment per atom for the dust
grains. We will assume that the emission detected by ATCA is
originated by spinning PAHs. The motivation for this assumption is the
excellent correlation between the radio emission and the 8\,$\mu$m map
in this region
\citep{cas08,Arce-Tord2020MNRAS.495.3482A}.

In order to fit the ATCA data using the SPDUST code, we fixed some of
the parameters that are well constrained in the literature for this
region. \citet{Habart2003} modeled the mid-IR line emission using a
PDR code and derived physical parameters for the $\rho$\,Oph\,W
filament. Using their results, we fixed the gas temperature and the
intensity of the radiation field. For the ionized Hydrogen and Carbon
abundances, we took the idealized values for PDRs that are listed in
\citet{dl98b}. We then fitted the SEDs using only 3 free parameters:
gas density (n$_H$), average dipole ($\beta$) and an additional
parameter of the grain size distribution ($a_{\rm cutoff}$) that
represents the minimum PAH size that is present in the region. This
last parameter is necessary to avoid shifting the spinning dust peak
to frequencies higher than $\sim30$\,GHz, as predicted by spinning
dust models for an ISM dust distribution in dense conditions such as
in this PDR. We note that some of the parameters in SPDUST are
expected to be highly correlated, for example the gas density,
temperature and radiation field. We avoid these degeneracies by fixing
most of the parameters to the physical conditions already inferred for
this region.

\begin{figure}
  \begin{center}
     \includegraphics[angle=90,width=0.49\textwidth]{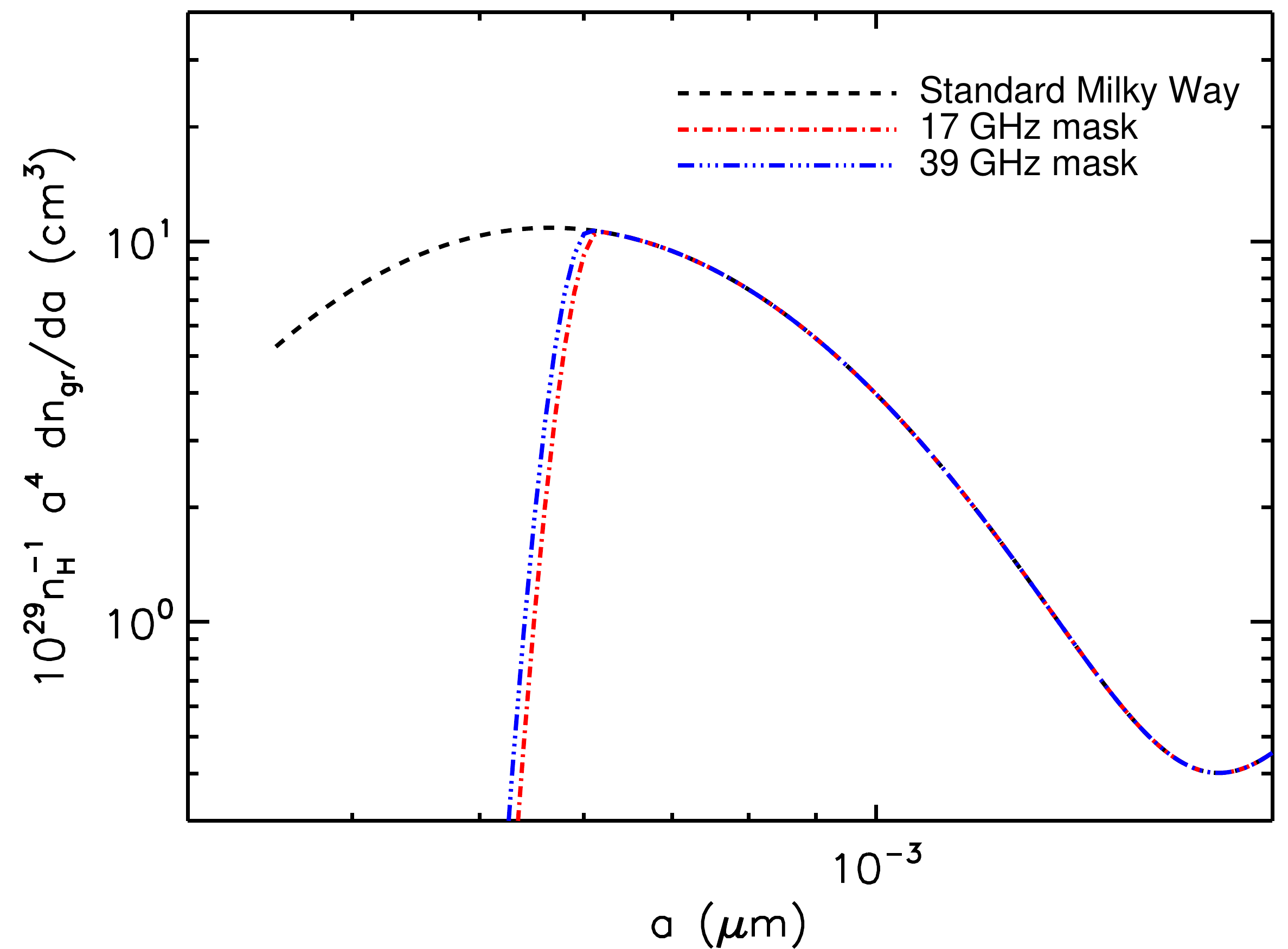}
  \end{center}
  \caption{Grain size distribution for sizes around 1\,nm for 3
    cases. In black is the prescription from \citet{Weingartner2001}
    for typical Milky Way parameters. In red and blue are the
    distributions arising from the spinning dust fit in the two
    studied regions. In these cases, a cutoff is needed in order to
    fit the SEDs in Fig. \ref{fig:sed}.}
  \label{fig:gsd}
\end{figure}

In SPDUST, the grain size distribution is parameterised as in
\citet{Weingartner2001}, where the contribution from PAHs is
characterized by two log-normal distributions. A typical curve using
standard parameters for the Milky Way is shown in black in
Fig. \ref{fig:gsd}. We introduced the $a_{\rm cutoff}$ parameter in
order  to adjust the region of the grain size distribution
that is most relevant to the spinning dust emission: the population of
the smallest grains. This additional parameter $a_{\rm cutoff}$
corresponds to a characteristic size below which we apply an
exponential cutoff modulating the size distribution, so effectively
defining a minimum size for the PAHs.

The data in the SEDs were fitted using the IDL routine {\tt mpfitfun}
\citep{MPFIT_2009}, that uses the Levenberg-Marquardt least-squares
fit to a function. We performed the fit using the SPDUST model for the
two regions shown in Fig. \ref{fig:rgb_mask}. The result from this
initial fit gave very similar values between the two regions for
$n_{\rm H}$ ($3.0\pm1.5$ vs $3.1\pm2.7$) and $\beta$ ($38.7\pm5.9$ vs
$35.1\pm8.4$). We thus decided to fix $n_{\rm H}$ and $\beta$ and only
fit for $a_{\rm cutoff}$.  Fig. \ref{fig:sed}  compares the best
fit SPDUST2 model curves with the SED data points.

The result of our fits are summarised in
Table\,\ref{tab:fit_params}. The key parameter to account for the
observed maximum at 30\,GHz is $a_{\rm cutoff}$, without which the
peak would shift towards $\sim$90\,GHz if fixing the physical
conditions to those determined independently in this PDR by
\citet{Habart2003}.  The difference in the free parameter, $a_{\rm
  cutoff}$, between the two regions is only 1.8\,$\sigma$ but it seems
to go in the direction expected in a PDR.  The minimum grain size
$a_{\rm cutoff}$ is slightly larger in the fit of the 17\,GHz mask
data. This means that in this region, there is a slightly lower
abundance of the smallest grains, compared to the other region.  This
behaviour is in agreement with intuition, as the 39\,GHz mask is more
exposed to the radiation from HD\,147889, which can result in a larger
number of the smallest PAHs due to the fragmentation of larger
ones. This provides a possible interpretation for the strong
morphological differences with frequency, which is reflected in the
local SEDs. \citet{Vidal2020MNRAS.495.1122V} recently concluded that
variations in the grain size distribution are also needed to explain
spinning dust morphology in the translucent cloud LDN\,1780.

The difference in $a_{\rm cutoff}$ between the two regions used to
link the multi-frequency morphological variations with spectral trends
may seem small. However, it is consistent with equipartition of
rotational energy \citep[e.g. Eq.\,13 and Eq.\,2
  in][respectively]{dl98b,Dickinson2018NewAR..80....1D}, which
suggests that a reduction in grain size from 6.3\,nm to 6.0\,nm would
shift equipartition rotation frequencies $\tilde{\nu}$ from 30\,GHz to
31.5\,GHz. The spectrum will be shifted accordingly, since for a
Boltzmann distribution of rotation frequencies, the emergent emissivity
$j_\nu/n_H$ is modulated by a high frequency Boltzmann cutoff
$\exp\left[-\frac{3}{2} \nu^2 / \tilde{\nu}^2 \right]$ \citep[see
  Eq.\,63 in][]{dl98b}.

Most of the SPDUST2 parameters were kept fixed in the optimization
summarised in Table\,\ref{tab:fit_params}. Yet some of these
parameters are expected to vary with depth into the PDR, and most
particularly the radiation field. Variations in $\chi$ may also play a
role in the spectral variations between the two SED extractions. We
tested for the impact of such variations by optimizing a model for the
SED in the 17\,GHz mask in which we decreased the UV field from our
default value of $\chi$=400, to $\chi=100$. The result was a slightly
better fit, with reduced $\chi^2_r=1.5$, and $a_{\rm{cutoff}} = 6.2
\pm 0.5$\r{A}. Therefore, even with a $\times$4 variation in the
intensity of the UV field, an increasing $a_{\rm{cutoff}}$ deeper into
the PDR seems to be a robust prediction of the SPDUST2 models.

\begin{table}
\centering
\caption{SPDUST2 fit parameters. Parameters without uncertainty were fixed to the ones reported in \citet{Habart2003}.}
\begin{tabular}{lHH}
\toprule
\textit{Parameter}
   &  \multicolumn{1}{c}{Mask$_{\rm{17\,GHz}}$}  
   &  \multicolumn{1}{c}{Mask$_{\rm{39\,GHz}}$}\\
     \midrule
                                                                     $n_{\rm{H}}$ [10$^3$ cm$^{-3}$] &\multicolumn{2}{c}{     3.2}      \\
                                                                                               T [K] &\multicolumn{2}{c}{   300.0} \\
                                                                                              $\chi$ &\multicolumn{2}{c}{   400.0} \\
                                                   $x_{\rm{H}} \equiv n_{\rm{H^+}}/n_{\rm{H}}$ [ppm] &\multicolumn{2}{c}{  1200.0}  \\
                                                   $x_{\rm{C}} \equiv n_{\rm{C^+}}/n_{\rm{H}}$ [ppm] &\multicolumn{2}{c}{   300.0}  \\
                                                                  $y \equiv 2n(\rm{H_2})/n_{\rm{H}}$ &\multicolumn{2}{c}{     0.0}   \\
                                                                                             $\beta$ &\multicolumn{2}{c}{    35.2}  \\
                                                                           a$_{\rm{cutoff}}$ [\r{A}] &    6.17 ,     0.04 &     6.07 ,     0.04 \\
          $\chi^2_r$ &\multicolumn{1}{c}{     2.9} & \multicolumn{1}{c}{     2.4} \\
\bottomrule
 \end{tabular}
\label{tab:fit_params}
\end{table}

\begin{figure}
  \begin{center}
     \includegraphics[angle=0,width=0.49\textwidth]{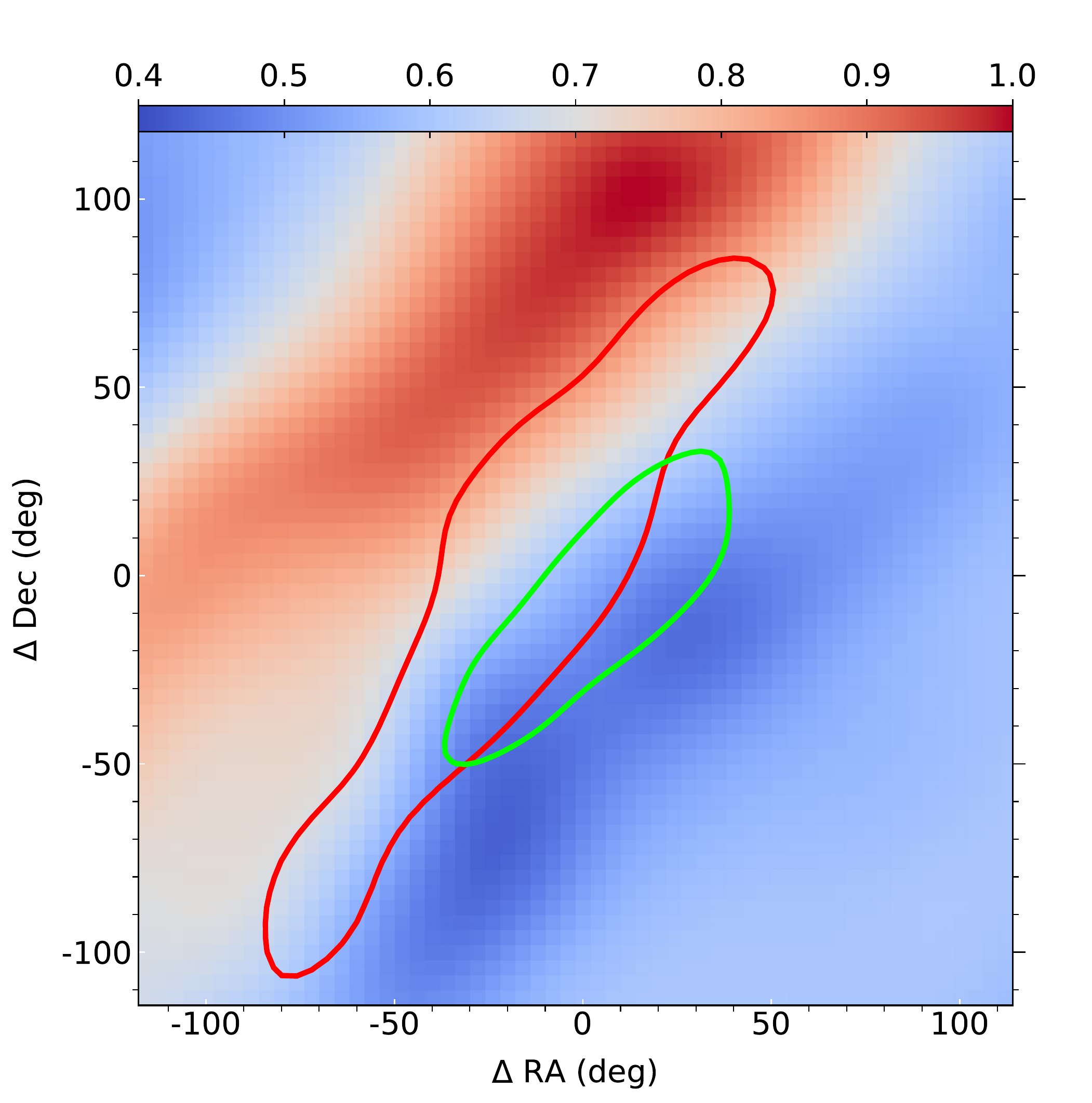}
  \end{center}
  \caption{Comparison between the ATCA continuum and a proxy for PAH
    size. We show an overlay of a 17\,GHz contour (in red,
    at 85\%-peak) and a 39\,GHz contour (in green, at 80\%-peak), on the ratio
    of {\em WISE} 12$\mu$m to 3.4$\mu$m. The increasing value of the ratio
    deeper into the PDR is consistent with the larger PAHs.}
  \label{fig:12_3}
\end{figure}

Further support for an increasing PAH size deeper into the PDR can be
found in a comparison with the {\em WISE} bands centred on 12\,$\mu$m
and 3.4$\mu$m, which each correspond to PAH bands and whose ratio is a
proxy for PAH size \citep{allamandola+85, ricca+12, croiset+16}.  The
relatively coarse angular resolution of the {\em WISE} images
(6.1\arcsec at 3.4$\mu$m and 15\arcsec at 12$\mu$m), compared to
IRAC\,8$\mu$m (2.5\arcsec), prevents their filtering for the ATCA+{\tt
  skymem} response. But we can nonetheless degrade the {\em WISE}
images to the coarsest ATCA beam (at 17\,GHz) for a multi-frequency
comparison. The smoothed images are not exactly comparable to the ATCA
mosaics, since we have not filtered for the ATCA response. But we hope
that any resulting bias in the following analysis is small, since our
synthesis imaging strategy corrects for missing ATCA antenna spacings
using a prior image in {\tt skymem}, and the main source of PSF
sidelobes is due to flux loss from missing antenna spacings at the
center of the $uv$-plane. We used the {\em WISE} images postprocessed
as in \citep[][]{Arce-Tord2020MNRAS.495.3482A} to produce
Fig.\,\ref{fig:12_3}, which illustrates that the gradient in peak
frequency across the filament is coincident with an increasing {\em
  WISE} 12\,$\mu$m/3.4$\mu$m ratio.  We quantify this trend using a
standard Pearson correlation test $r$ \citep[e.g. same as $r_{\rm
    sky}$ in Eq.\,11 of][]{Arce-Tord2020MNRAS.495.3482A}, so similar
to $r_w$ in Eq.\,\ref{eq:rw} but without the weights, and instead
adjusting the field of extraction to avoid the noise at the edge of
the ATCA mosaics. The resulting Pearson $r$ are listed in
Table\,\ref{table:r_WISE}. We recover the same trend as in
Table\,\ref{table:xcorr}, both $r$ and $r_w$ point at 20\,GHz as the
best match to IRAC\,8$\mu$m. However, the ATCA map that best traces
the shorter {\em WISE} wavelength is 39\,GHz. The excellent
correlation between ATCA 33/39\,GHz with  3.4\,$\mu$m and also
between ATCA 20/17\,GHz with the 8\,$\mu$m template confirm the strong
correlation between AME and PAH emission in this region.

The AME-PAH connection was put in doubt by \citet{Hensley2016} based
on a full-sky analysis on angular scales of 1\,deg. Indeed, when taken
as a whole, the $\rho$\,Oph cloud is a good example of the breakdown
of the correlation between PAHs tracers and AME, since S\,1, the
brightest nebula in the complex in IRAC\,8$\mu$m and also in {\em
  Spitzer}-IRS 11.3$\mu$m PAH band \citep[][their Table\,2]{cas08},
has no detectable EME signal. However, it appears that EME and PAHs do
correlate very tightly in higher angular resolution observations, and  in
regions where EME is present. Another example of excellent
correspondence between AME and PAH emission is shown clearly in
LDN1246 \citep{Scaife2010}.

The predictions for the grain-size distribution reported here should
also be compared to the IR spectra available for $\rho$\,Oph\,W. Under
the spinning dust hypothesis for EME, a complete model should
reproduce simultaneously the radio SED as well as the IR spectra,
which are both due to the same underlying dust population. Here we
limit the scope of this report on the new ATCA observations to only
the radio part, highlighting the need for a future modeling effort.

\begin{table}
\centering
\caption{ATCA - {\em WISE} correlation statistics}
\label{table:r_WISE}
\begin{tabular}{lcc}
\hline 
Frequency & {\em WISE}\,3.4$\mu$m   &  IRAC\,8$\mu$m    \\
MHz    &                          & \\ \hline
     17481 &   0.01$^a$  & 0.82   \\   
     20160 &   0.24  & 0.90   \\   
     33157 &   0.91  & 0.68   \\   
     39157 &   0.93  & 0.44   \\ \hline
\end{tabular}
\begin{flushleft}
  $^a$ Pearson $r$ coefficients. All values bear a 1$\sigma$  uncertainty of 0.04.
  \end{flushleft}
\end{table}

\section{Carbon RRL search} \label{sec:crrls}

The main spin-up mechanisms that could lead to VSG rotation
frequencies of up to $\sim$30\,GHz may be either radiative torques or
plasma drag \citep[][]{dl98b}. Interestingly, the brightest near-IR
nebulae in $\rho$\,Oph, i.e. S\,1 and SR\,3, have no radio counterparts
at cm-wavelengths. Yet the circumstellar environments of embedded
early-type stars correspond to the highest UV-radiation intensities.
The absence of radio sources coincident with the IR-bright
circumstellar dust about S\,1 and SR\,3 cannot be reconciled with VSG
depletion, as marginally shown by the CBI observations reported by
\citet{cas08}, and confirmed with CBI2 in \citet[][]{Arce-Tord2020MNRAS.495.3482A}.
Radiative torques seem unlikely to explain the strong radio signal
from the $\rho$\,Oph\,W filament.

The alternative source of rotational excitation, plasma drag, is due
to the interaction of the grain dipoles with passing ions - namely
H$^+$ or C$^+$ in the context of PDRs. If so the spinning dust
emissivities would be best understood in terms of an emission measure:
$I_\mathrm{cm} \propto N_\mathrm{VSG} N(\mathrm{C}+)$.  An
observational test of the plasma-drag hypothesis requires measurements
of the C$^+$ abundance, as can be inferred using radio carbon
recombination lines. The faint or absent  cm-wavelength signal
from the circumstellar nebulae around S\,1 and SR\,3 in the radio maps
may be due to these stars being too cold to create conspicuous C\,{\sc
  ii} regions \citep[][]{cas08}.

\citet{pan78} examined the most complete set of RRL data towards
$\rho$\,Oph to date. The line profiles observed at lower frequencies
have widths of 1.5\,km\,s$^{-1}$ FWHM.  They mapped the neighbourhood
of S\,1, but did not extend their coverage to $\rho$\,Oph\,W,
unfortunately. The highest frequency RRLs considered by \citet{pan78}
are C90$\alpha$ and C91$\alpha$, at $\sim$9\,GHz, which they
interpreted as stemming from circumstellar gas about S\,1, with
electron densities $n_{\rm e} \sim 15\,$cm$^{-3}$ and $T_{\rm e} \sim
150\,$K. This circumstellar C\,{\sc ii} region was inferred to be less
than $\sim$2\,arcmin in diameter, and surrounded by a diffuse halo
with $n_{\rm e} \sim 1\,$cm$^{-3}$, traced by the lower frequency
carbon RRLs.


We searched for carbon RRLs in the ATCA+CABB data, with a 2\,GHz
bandwidth centred on 17481\,MHz. Three $\alpha$-type RRLs fall into the
17481\,GHz IF: C71$\alpha$ 18.00153, C72$\alpha$ 17.26682, C73$\alpha$
16.57156. No carbon RRLs are detected near the systemic velocity of
$\rho$\,Oph\,W \citep[which is $V_\mathrm{lsr} =
  +3\,$km\,s$^{-1}$][]{1974ApJ...189..253B,pan78}.  In the 17481\,GHz
IF the velocity width of each channel is $\sim$16\,km\,s$^{-1}$. The
noise in single-channel reconstructions is $\sim$2\,mJy\,beam$^{-1}$,
for a 30\,arcsec beam FWHM. Assuming that the line is unresolved and
is diluted in such broad channels, this upper limit is a factor of two
looser than that obtained by \citet{cas08} using Mopra.

For a rough estimate of carbon RRL intensities in $\rho$\,Oph\,W, we
take a depth of 0.04\,pc, which at a distance of 135\,pc subtends
1\,arcmin, an ionisation fraction of $10^{-4}$, due to carbon
photoionisation, $T_{\rm e} = 100\,$K, and a H-nucleus density of
$n_\mathrm{H} = 10^5$\,cm$^{-3}$ \citep[these values are similar to
  those reported previously for $\rho$\,Oph\,W,
  e.g.][]{Habart2003}. The peak intensity of the emergent C71$\alpha$
is 18\,mJy\,beam$^{-1}$, for LTE\footnote{The LTE deviations become
important for $n=72$ at $T_{\rm e}< 100\,$K, i.e. the $n=72$
population departure coefficient relative to LTE is $b < 1$ and the emergent intensities are proportionally fainter}, with a
1.5\,km\,s$^{-1}$ FWHM, and a 30\,arcsec beam. When diluted in the
$\sim$\,16\,km\,s$^{-1}$ channels of CABB, the expected signal drops
down to $\sim$\,1\,mJy\,beam$^{-1}$, or close to the limits obtained
with Mopra. However, the expected CRRLs intensities should be within
easy reach with the Atacama Large Millimetre Array (ALMA), as long as
the spectral resolution is not degraded much beyond
$\sim$0.5\,km\,s$^{-1}$. The spectra line data could be acquired as
part of future observations to map the EME signal in $\rho$\,Oph\,W at
$\sim$40\,GHz with the Band\,1 receivers currently under construction,
and which should yield a noise level of 3\,mJy\,beam$^{-1}$ in 40\,min
and in 0.5\,km\,s$^{-1}$ channels.

\section{Conclusion} \label{sec:conc}

ATCA+CABB multi-configuration mosaics of the $\rho$\,Oph\,W PDR
resolve the filament with $\sim$30\,arcsec resolutions from 5\,GHz to
39\,GHz.  Since the signal fills the primary beam a special purpose
imaging synthesis strategy ({\tt skymem}) was applied to compensate
for flux loss and mitigate sidelobe oscillations with the
incorporation of an image prior.

The multi-frequency 17\,GHz to 39\,GHz mosaics reveal spectral
variations within $\rho$\,Oph\,W. The radio signal follows the near-IR
filament, but it is progressively shifted towards the UV source at
higher frequencies. Such morphological differences in frequency
reflect changes in the radio spectrum as a function of position in the
sky.  While the morphological trends with frequency are qualitative,
the corresponding spectral variations in terms of the SEDs are not
significant given the systematic uncertainties.


The SED of $\rho$\,Oph\,W, with a very narrow peak at $\sim$30\,GHz,
is reminiscent of spinning-dust. The physical conditions inferred
under this hypothesis, using an optimization of selected  free-parameters in
the SPDUST package, are consistent with those derived in the
literature, but require a minimum grain size cutoff and relatively
large electric dipoles. The cutoff in the grain sizes is particularly
well constrained as a standard ISM size distribution would shift the
peak of the spectrum towards $\sim$90\,GHz. The spinning dust model
accounts for the measured intensities, and suggests that the
qualitative morphological differences can be interpreted in terms of
an increasing minimum grain size deeper into the PDR. 

Further sampling of the spinning dust spectrum in $\rho$\,Oph\,W at
$\sim$50GHz with ALMA, in the context of the data reported here, would
provide strong constraints on the minimum PAH size. Eventually, the
predictions obtained from the rotational emission of PAHs should be
tested against a physical model for the IR PAH bands in
$\rho$\,Oph\,W.

\section*{Acknowledgments}

We thank the referee, Yvette Chanel Perrott, who provided important
input for the presentation of the {\tt skymem} algorithm and for the
interpretation of the SED fits, in addition to constructive comments
on the analysis  and a thorough reading. We also acknowledge
interesting discussions and comments from Kieran Cleary, Roberta
Paladini, Jacques Le Bourlot and Evelyne Roueff.  S.C. acknowledges
support from a Marie Curie International Incoming Fellowship
(REA-236176) and by FONDECYT grant 1171624.  MV acknowledges support
from FONDECYT through grant 11191205. GJW gratefully thanks the
Leverhulme Trust for the award of an Emeritus Fellowship.

\section*{Data Availability}

The {\tt skymem} package can be found at \url{https://github.com/simoncasassus/SkyMEM}. The corresponding author
will provide help to researchers interested in porting {\tt skymem} to
other applications.  The {\tt skymem} code repository also includes, as
an example application, the sky-plane version of the data underlying
this article. The unprocessed visibility dataset can be downloaded
from the Australia Telescope Online Archive at \url{https://atoa.atnf.csiro.au/}.  The corresponding author will share
the calibrated visibility data on reasonable request.



%
%

\input{report_ATCA_ROPHW.bbl}

\bibliography{merged_EME.bib}



\appendix

\section[]{Image reconstruction} \label{sec:appendix}



\begin{figure*}
  \begin{center}
    \includegraphics[width=0.95\textwidth,height=!]{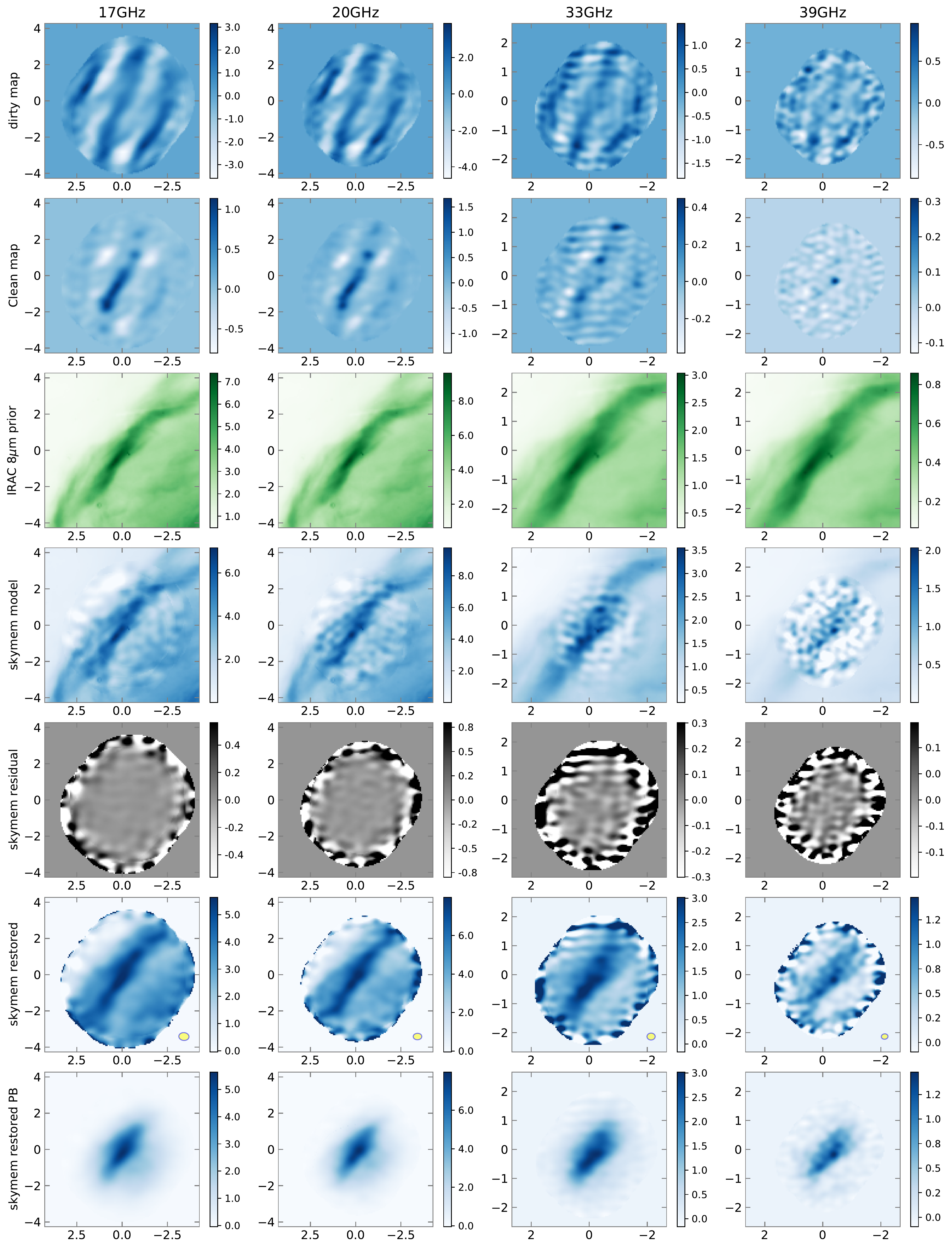}
  \end{center}
  \caption{\label{fig:skymem} Synthesis imaging of the ATCA
    observations of $\rho$\,Oph\,W. $x-$ and $y-$ axis correspond to
    offset R.A. and Dec. in arcmin. This
    panel of images is organized as a table, where each image
    corresponds to the frequency given in  column headers, for the
    synthesis imaging schemes given in  line headers. We have, from
    top to bottom, the dirty and clean mosaics calculated with
    Miriad, the prior image scaled to each frequency, followed by 
     the {\tt skymem} model image, its associated residual mosaic,
    and the restored mosaic, shown also  also after multiplication
    by the mosaic attenuation  (labeled `skymem restored PB'). The beam ellipses
    are shown in the restored images. The colour units are mJy\,beam$^{-1}$. These images have not been point-source
    subtracted.  }
\end{figure*}

The traditional image-reconstruction algorithm Clean is not ideal for
extended sources that fill the beam, especially with sparse
$uv$-coverage.  Initial trials at imaging using the Miriad task
`clean' resulted in large residuals, with an intensity amplitude much
greater than that expected from thermal noise, and with a spatial
structure reflecting the convolution of the negative synthetic
side-lobes with the morphology of the source (see
Fig.\,\ref{fig:skymem}). Attempts to improve dynamic range using the
`maxen' task in Miriad gave worse results.  We therefore designed a
special-purpose image reconstruction algorithm, based on sky-plane
deconvolution (hereafter {\tt skymem}), and that allows the
incorporation of priors to recover the larger angular scales.

The present sky-plane approach is an alternative to similar
non-parametric imaging synthesis strategies based on a $uv$-plane
approach, in which model visibilities are compared to the
interferometer data. An example package for such $uv$-plane approaches
is {\tt uvmem} \citep[][]{Casassus2006ApJ...639..951C,
  Carcamo2018A&C....22...16C}, which has been applied to diffuse ISM
data \citep[such as the CBI and CBI2 observations of $\rho$\,Oph,
][]{cas08, Arce-Tord2020MNRAS.495.3482A} as well as in compact sources
\citep[e.g. such as VLA and ALMA observations of protoplanetary
  discs][]{ Casassus2018MNRAS.477.5104C, Casassus2019MNRAS.483.3278C,
  Perez2019AJ....158...15P}. Both the sky-plane and the $uv$-plane approaches
should of course be equivalent, but in the sky plane we avoid delicate
issues with visibility gridding. In this application of {\tt skymem}
we rely entirely on the Miriad gridding machinery\footnote{the python version available on {\tt github} is being integrated with the CASA framework}.

In a sky-plane formulation of image synthesis, the data correspond to
the dirty maps $\{ I^D_j \}_{j=1}^f$ for each of the $f$ fields in the
mosaic.  In order to obtain a model sky image that fits the data we
need to solve the usual deconvolution problem, i.e. obtain the model
image $I^m$ that minimizes a merit function $\mathcal{L}$:
\begin{equation}
  \mathcal{L} = \chi^2 - \lambda \mathcal{S}(I^m), \label{eq:L}
\end{equation}
where
\begin{equation}
  \chi^2 = \sum_{j=1}^f  \sum_{i=1}^n  w_j(\vec{x}_i) \left( I^D_i(\vec{x}_i) -  I^{Dm}_j(\vec{x}_i) \right)^2.
\end{equation}
The sums extend over the number of fields, $f$, and over the number of
pixels in the model image, $n$. Each of the model dirty maps,
$I^{Dm}_j $, correspond to the convolution of the attenuated $I^{m}$
with the synthetic beam $B_j$,
\begin{equation}
I^{Dm}_j = (I^{m}_j A_j)  \ast B_j, \label{eq:IDm}
\end{equation}
where $\{ A_j \}_{j=1}^f$ are the primary beam attenuations for all
fields. The weight
map for a field $j$ is given by $w_j = 1 / \bar{\sigma}_j^2$, where
$\bar{\sigma}_j$ is the theoretical noise map as calculated with the `sensitivity'
option to the task `invert' in Miriad. This noise map is simply the thermal
noise expected in the dirty map divided by the primary beam
attenuation.


After several trials with a variety of functional forms for
$\mathcal{S}$, we found that we obtained best results by simply using
$\lambda = 0$, i.e.  with pure $\chi^2$ reconstructions. Image
positivity of its own provided enough regularization. We used the Perl
Data Language (PDL) for high-level data processing, and the
optimization was carried out with a PDL-C patch to the Fletcher-Reeves
algorithm in its GSL implementation. We enforced positivity by
clipping $I^m>0$ at each evaluation of $\mathcal{L}$ and its
gradient. A couple of aspects of the implementation of {\tt skymem}
are worth mentioning. In the convolution of Eq.\,\ref{eq:IDm} the
kernel should not be normalized, as would be the case for
smoothing. Instead, to yield $I^{Dm}_j$ in Jy\,beam$^{-1}$ units,
$B_j$ should be scaled by the number of pixels in a beam $\Omega_G /
(\delta x)^2 $, where $\Omega_G$ is the clean beam solid angle (see
below) and $\delta x$ is the pixel scale. Another relevant aspect is
the evaluation of the gradient of $\chi^2$, which can be written
\begin{equation}
  \frac{\partial \chi^2 }{\partial I^m(\vec{x}_i) } =
  \sum_{j=1}^f  \frac{2 A_j(\vec{x}_i)}{\bar{\sigma}^2} \left[ \left(I^{Dm} - I^D\right)  \ast B_j  \right]_{\vec{x}_i}.
\end{equation}

The initial condition is important for the optimization as its
parameter space is very structured. A blank initial image performed
better that Clean, but yet lower values of $\mathcal{L}$ were obtained
by starting with an image known to approximate the radio signal, which
we call the prior image.  The initial image we chose is the
IRAC\,8$\mu$m map, in its original angular resolution but filtered for
point sources. This version of the IRAC\,8$\mu$m map was multiplied by
a representative dimensionless radio/IR correlation slope of
$5.2\times10^{-4}$ \citep[extrapolated from the slopes reported
  in][]{Castellanos2011MNRAS.411.1137C}, so that the flux densities
fall within the order of magnitude of the observed CBI flux densities
in such sources.  We then refined the intensity scale to exactly match
that of the the ATCA observations in the following way.  We simulated
ATCA observation on the IRAC\,8\,$\mu$m template, with identical
$uv$-plane coverage as the  observations, and calculated the dirty
maps $\{ I^D_{\mathrm{IRAC}j} \}_{j=1}^f$ for each pointing $j$ in
these mock data. The best fit correlation slopes $\{ s_j \}_{j=1}^f$,
defined by $I^D_j = s_j I^D_{\mathrm{IRAC}j}$, are each given by
\begin{equation}
  s_j = \frac{\sum_i w_j(\vec{x}_i)
    I^D_{\mathrm{IRAC}j}(\vec{x}_i) I^D_j(\vec{x}_i)} {\sum_i
    w_j(\vec{x}_i) (I^D_{\mathrm{IRAC}j}(\vec{x}_i))^2 }.
\end{equation}
Finally, the prior image corresponds to this IRAC\,8\,$\mu$m template
scaled by $\langle s \rangle$, the mean correlation slope taken over
all pointings.  These prior images and their associated intensity
scales are shown in
Fig.\,\ref{fig:skymem}. Table\,\ref{table:sfactors} lists the values
for $\langle s \rangle$ and $\sigma(s)$ at each frequency.  It is
interesting to compare with the radio-IR correlation slopes $a(\nu)$
listed in Table\,\ref{table:xcorr}. The larger dispersion of $s(\nu)$
with increasing frequency could reflect either real spectral
changes. For the {\tt skymem} simulations on the IRAC template, all
$s(\nu)\equiv 1$ .


%
\begin{table}
\centering
\caption{Scale factors for the {\tt skymem}  priors}
\label{table:sfactors}
\begin{tabular}{lll}
  \hline
  
Frequency$^a$ & $\langle s  \rangle ^b$ &  $\sigma(s)\,^c$   \\ \hline
5500 &  0.015  &  -  \\ 
8800 &  0.114  &  -  \\ 
17481 &  1.694  &  0.163  \\ 
20160 &  3.162  &  0.269  \\ 
33157 &  2.495  &  0.546  \\ 
39157 &  1.090  &  1.021  \\ \hline
\end{tabular}
\begin{flushleft}
$^{a}$ Centre frequency in MHz.
$^{b}$ Average and $^c$ dispersion taken over all pointings. 
\end{flushleft}
\end{table}

The resulting model images are shown in Fig.\,\ref{fig:skymem}. It can
be appreciated that the free parameters in the model image are
modified relative to the input prior only within the field of the ATCA
mosaic. It is also interesting to note that the low spatial
frequencies of the prior are preserved, since the ATCA data provide no
information that would constrain them.

Image restoration was obtained by smoothing the model image with the
clean beam\footnote{which is an elliptical Gaussian} $G$ in a
reference field (that also sets the Jy\,beam$^{-1}$ units), and by
adding the linear mosaic of dirty residuals $R^D$,
\begin{equation}
  I^R = I^m \ast G + R^D.
\end{equation}
The residual image  for each pointing $j$ is $T^D_j= I^D_j  - I^{Dm}_j$, so 
\begin{equation}
  R^D = \frac{ \sum_j  w_j T^D_j  A_j       }{ \sum_j w_j A_j^2}. \label{eq:mosaic}
\end{equation}
The residual and restored images are shown in
Fig.\,\ref{fig:skymem}. The residuals are adequately thermal, but the
linear mosaic generated with the formula in Eq.\,\ref{eq:mosaic}
amplifies the noise at the edges of the field. Thus we also provide in
Fig.\,\ref{fig:skymem} a version of each restored images after
multiplication by the mosaic attenuation pattern $\mathcal{A}$ to highlight the
regions with smallest thermal errors, with 
\begin{equation}
  \mathcal{A}(\vec{x}) = \frac{\bar{\sigma}_\circ }{\bar{\sigma}_R(\vec{x})},
\end{equation}
where $\bar{\sigma}_\circ$ is the minimum value in the theoretical noise image,
\begin{equation}
  \bar{\sigma}_R = \sqrt{  \frac{1}{ \sum_j w_j A_j^2}}. \label{eq:noisemosaictheor}
\end{equation}

The dynamic range of the resulting {\tt skymem} images can be
estimated by calculating the mean and dispersion of the residuals, i.e.
\begin{equation}
  \langle R^D \rangle = \frac{\sum_j \bar{w}_R(\vec{x}_j)  R^D(\vec{x}_j) }{ \sum_j \bar{w}_R(\vec{x}_j)},
\end{equation}
and
\begin{equation}
  \sigma_\circ = \sqrt{ \frac{\sum_j \bar{w}_R(\vec{x}_j) \left(
      R^D(\vec{x}_j) - \langle R^D \rangle \right)^2} { \sum_j
      \bar{w}_R(\vec{x}_j)}}.
\end{equation}
The values for $\sigma_\circ$ are given in Table\,\ref{table:noise},
where we see that they come close to the theoretical ATCA
sensitivity. The noise image measured using the residuals rather than the
theoretical sensitivy can be written as
\begin{equation}
  \sigma_R = \sigma_\circ  \frac{ \bar{\sigma}_R}{\bar{\sigma}_\circ} . \label{eq:noisemosaic}
\end{equation}

\begin{table}
\centering
\caption{Dispersion of residuals and expected theoretical noise, in $\mu$Jy\,beam$^{-1}$}
\label{table:noise}
\begin{tabular}{lll}
  \hline
  Frequency$^a$ & $\sigma_\circ^b$  & $\sigma_{\rm ETC}^c$  \\ \hline
  5500  & 24  &  5   \\
  8800  & 23  &  6   \\
  17481 & 13  & 12    \\
  20160 & 25  & 19    \\
  33157 & 38  & 17   \\
  39157 & 23  & 16   \\
 \hline
\end{tabular}
\begin{flushleft}
  $^{a}$ Centre frequency in MHz.
  $^{b}$ Measured dispersion of {\tt skymem} residuals.
  $^{c}$ Expected noise level in full scans, from the Exposure Time Calculator at  \url{https://www.narrabri.atnf.csiro.au/myatca/interactive_senscalc.html}.
\end{flushleft}
\end{table}



%
%
%

\bsp	
\label{lastpage}
\end{document}